\documentclass{rspublic}

\usepackage[dvips]{color}       
\usepackage{graphics}           
\usepackage{bm}                 
\usepackage{amsmath}            
\usepackage{amsfonts}
\usepackage{amssymb}
\usepackage{latexsym}           

\usepackage[round,colon,authoryear]{natbib}
\bibpunct[, ]{(}{)}{;}{a}{}{,}
\newcommand{\identity}{\leavevmode\hbox{\small1\kern-3.2pt\normalsize1}}

\begin{document}

\title
	[Quantum-assisted biomolecular modelling]
	{Quantum-assisted biomolecular modelling}

\author
       [S.~Harris, V.~Kendon]
       {Sarah Harris and Vivien M.~Kendon}

\affiliation{School of Physics and Astronomy, University of Leeds, Leeds LS2 9JT, UK}

\label{firstpage}

\maketitle

\begin{abstract}{computational biophysics, quantum computation}
Our understanding of the physics of biological molecules,
such as proteins and DNA, is limited because the approximations
we usually apply to model inert materials are not in general
applicable to soft, chemically inhomogeneous systems.
The configurational complexity of biomolecules means the
entropic contribution to the free energy is a significant
factor in their behaviour, requiring detailed dynamical
calculations to fully evaluate.
Computer simulations capable of taking all interatomic
interactions into account are therefore vital.
However, 
even with the best current supercomputing facilities,
we are unable to capture enough of the most interesting
aspects of their behaviour to properly understand how they work.
This limits our ability to design new molecules, to treat diseases,
for example.  Progress in biomolecular simulation depends crucially
on increasing the computing power available.  Faster classical
computers are in the pipeline, but these provide only incremental
improvements.
Quantum computing offers the possibility of performing huge numbers
of calculations in parallel, when it becomes available.
We discuss the current open questions in biomolecular simulation,
how these might be addressed using quantum computation and speculate
on the future importance of quantum-assisted biomolecular modelling.
\end{abstract}

\section{Introduction}
\label{sec:intro}

The chemical complexity of biological macromolecules enables them to
perform extraordinary functions.  Biomolecular recognition, enzyme
catalysis, self-organisation, and molecular motors are central to all
cellular processes, but remain poorly understood theoretically.
This limits our ability to develop new drugs to inhibit or promote
a particular process, or to design our own nanoscale devices
with bespoke functions.
If we had an equivalent theoretical understanding of biological
systems as we have of semiconductors, then whole new regimes of
bio-inspired engineering at the nanoscale would become possible.

Experimental methods to investigate biomolecular structure at the
atomic level, such as X-ray crystallography and NMR (nuclear magnetic
resonance), have revolutionised our understanding of biomolecular function.
Computer simulation is also extremely important in the biomolecular sciences,
because it allows a physical model of the system to be constructed, but
does not require such severe approximations as phenomenological models.
Computer simulations at the atomistic level have proven enormously
beneficial in molecular biology; for example they are routinely used to
study molecular recognition and docking
(which is of importance in drug design) and are
integral to NMR structure refinement for biomolecules.
However, due to the computational expense of the calculations, we are only
able to use these methods to study small biomolecules (usually nm) for
short timescales (usually ns). This is a serious limitation; many of the
important conformational changes associated with biomolecular function
occur over far longer timescales ($\mu$s--ms), and many functional
biomolecular systems are large protein complexes ($\sim$50nm).

We are therefore looking in the first instance for improvements
in the length of time we can run for of $\sim 10^3$--$10^4$, and
in the systems size of $\sim 10$--$10^2$, a combined scaling of
$\sim 10^5$--$10^6$ -- a million times larger than current state-of-the-art.
Ultimately, we would like to simulate much larger systems, up to the size
of a whole cell, another million times larger, and beyond.
Undoubtedly, a good deal of the scaling up has to be done by refining the
models to be more efficient \citep[e.g., ][]{moritsugu09a}.
But without significantly more computing
power it will be difficult to advance our understanding
to the point where the models can deliver that level of improvement.

In this paper, we consider how future developments in
quantum computing could enable our biomolecular simulations to reach
new regimes.  High performance quantum computing (HPQC) has recently been
proposed in a fully scalable main-frame architecture based on
topologically encoded photonic qubits \citep{devitt08a}.
To fully exploit HPQC for biomolecular simulation will require significant
algorithm redesign to obtain a quantum mediated speed up over classical
methods.

\section{Biomolecules and their roles}
\label{sec:biom}

Biomolecules are polymers made of discrete building blocks that impart
functional specificity.  It is the enormous diversity and
versatility of these building blocks which enables biomolecules to perform
such a remarkable range of functions. Cells contain proteins, nucleic
acids (DNA and RNA), lipids, sugars and numerous other small organic
molecules which participate in cell regulation. The nucleic acids DNA and
RNA act as storage and messenger molecules for genetic information.
Proteins can act by transfering biological information, many are
catalysts of biochemical reactions, others are cellular scaffolds 
responsible for structural integrity in the cell, and some are
molecular machines which couple chemical energy to a mechanical process.
Organisation is vital in such a busy molecular environment.
Lipid membranes compartmentalise cells into regions that perform
specific functions.  Communication with the rest of the
cell is achieved by way of membrane proteins which act as switchable
pores. To understand biology at the molecular level, it is necessary to relate
the complex structure, the diverse chemistry and the
anharmonic dynamics of biomolecules to their specific function within the cell.

\subsection{Biomolecular structure and function}
\label{ssec:chem}

\begin{figure}
\begin{center}
\resizebox{1.0\columnwidth}{!}{\rotatebox{0}{\includegraphics{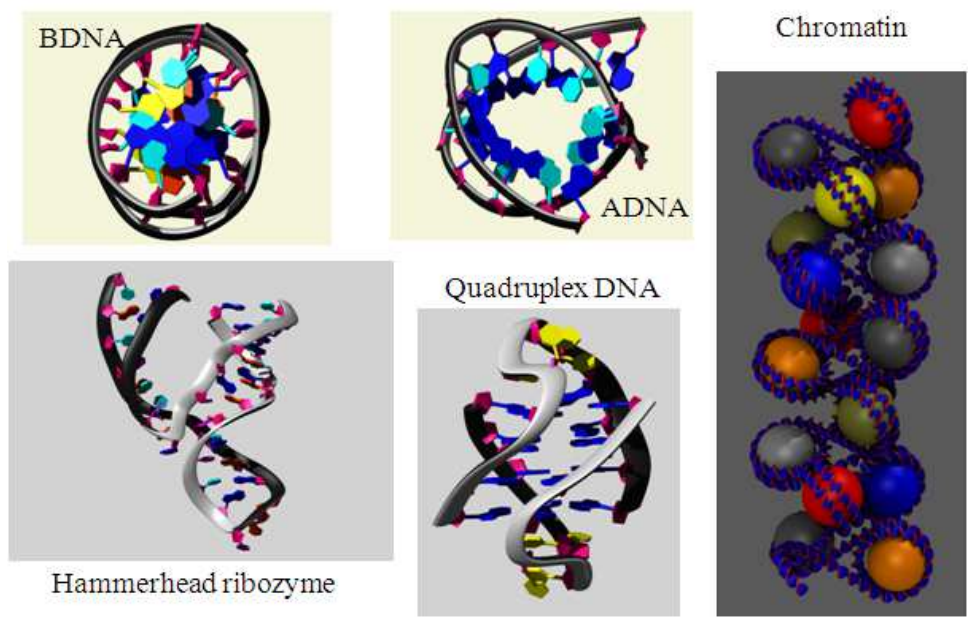}}}
\end{center}
\caption{Nucleic acids in Nature:
The most common DNA conformation is B-form DNA (top left),
A-DNA (top centre) occurs under dehydrated conditions,
and quadruplex DNA (bottom centre) is thought to be
formed at the ends of chromosomes.
Chromatin structure (right).
Hammerhead ribozyme (bottom left), an RNA enzyme.
Figures produced using Chimera \citep{Chimera} and VMD \citep{VMD}.}
\label{fig:DNA}
\end{figure}

Figure \ref{fig:DNA} shows a small but representative selection of the nucleic acid structures
that are of biological importance. DNA carries the genetic code through
the specific biological relationship between the sequence of the four
DNA bases adenine (A), guanine (G), thymine (T) and cytosine (C) and the
sequence of amino acids in a protein.
The most common form of DNA is known as B-form DNA or
canonical DNA (see figure \ref{fig:DNA} top left), although other forms,
such as A form DNA (figure \ref{fig:DNA} top centre) and quadruplex
(four-stranded) DNA (see figure \ref{fig:DNA} bottom centre), are also
important.  In eukaryotes, DNA is also associated with histone
proteins which compact the genetic material into a structure known as
chromatin so that it will fit into the nucleus (see figure \ref{fig:DNA}
right). One of the most important functional roles of RNA in the cell is
to act as a messenger molecule between DNA and proteins. However, some
viruses use RNA rather than DNA as their genetic material, and RNA
molecules such as the Hammerhead ribozyme (see figure \ref{fig:DNA}
bottom left) are also sufficiently chemically active that they can act
as enzymes.

\begin{figure}
\begin{center}
\resizebox{1.0\columnwidth}{!}{\rotatebox{0}{\includegraphics{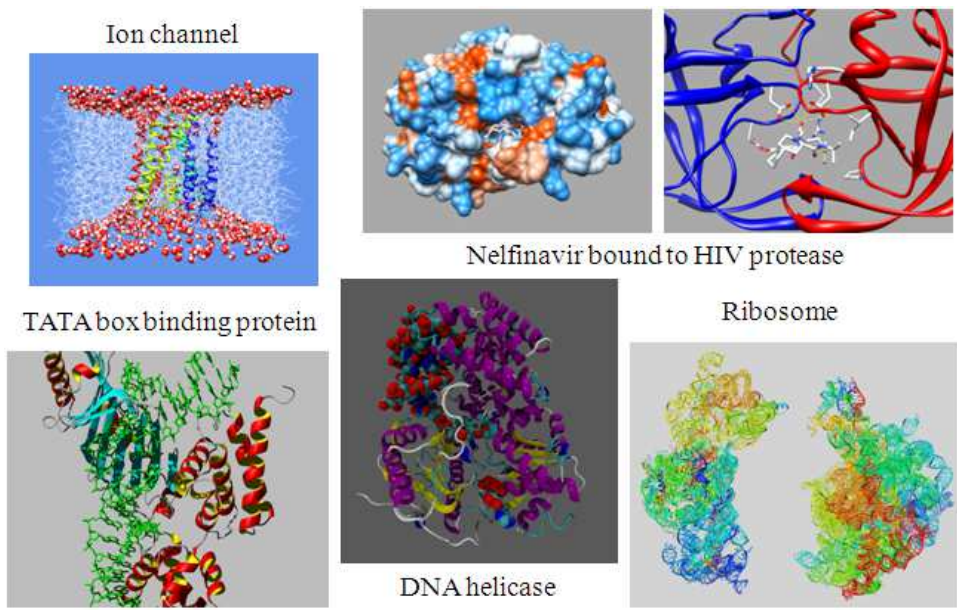}}}
\end{center}
\caption{Protein folding, recognition and machinery.
Structure of the ion channel (X) embedded in a lipid membrane (top left),
two representations of the inhibitor Nelfinavir bound
to HIV protease (top middle/right),
TATA box binding protein bound to DNA (bottom left),
DNA helicase molecular motor (bottom middle),
ribosome (bottom right).}
\label{fig:amino}
\end{figure}

Proteins are constructed by the ribosome (shown in figure \ref{fig:amino}
bottom right), which
catalyses the formation of a polymer chain made of the sequence of amino
acid residues encoded by the messenger RNA template.
Before the protein is biologically functional, it must fold into a
tightly packed globular structure.
Protein folding takes place over timescales of around $1ms$,
with even the fastest folders requiring more than $10 \mu s$,
see \cite{proteinfold}.
In the aqueous environment of the
cytoplasm, proteins often have a hydrophobic core surrounded by a
hydrophilic shell whereas membrane proteins, which are located in a
hydrophobic lipid environment, generally have a hydrophobic exterior and
a water filled hydrophilic core. An example of a membrane protein is
shown in figure \ref{fig:amino} (top left). For a more detailed
description of the biochemistry of nucleic acids and proteins,
see \cite{stryer}.

\subsection{Thermodynamics of biomolecules}
\label{ssec:thermo}

Both protein folding and molecular recognition are driven by the
thermodynamic requirement that the free energy is a minimum at equilibrium.
Even though the cell is full of dynamical non-equilibrium
processes, these take place on long enough timescales for the
biomolecules themselves to be in local thermal equilibrium.
A protein adopts its native structure because folding reduces the free energy.
The TATA box binding protein, shown in figure \ref{fig:amino} (bottom left),
acts as a switch to activate gene transcription.
It will preferentially bind to DNA containing the sequence TATA,
because this gives the most favorable change in free energy.

The changes in free energy that take place during protein folding
or molecular recognition are so subtle that there is currently no
theoretical method capable of either predicting the folded state of a
protein from a knowledge of its amino acid sequence, or of determining the
binding constant of a protein for its molecular target.
The free energy change $\Delta G$ is made up of two separate contributions:
 \begin{equation}
\Delta G = \Delta U - T\Delta S.
\end{equation}
The energetic contribution $\Delta U$ is made up firstly of
all of the favorable chemical interactions which promote the formation
of a folded protein or a biomolecular complex.
These are electrostatic interactions between oppositely charge amino acids
residues, or protein/nucleic acid interactions;
favorable van der Waals interactions and hydrogen bonds.
In molecular recognition, these interactions occur
due to shape and chemical complementarity of the reactants in a highly
specific manner, in much the same way as a key fits a particular lock.
An example is the interaction between the HIV protease inhibitor
Nelfinavir and its molecular target, as shown in figure
\ref{fig:amino} (top centre/right).
However, biomolecules are also inherently soft,
and often binding an external molecule induces a
conformational change that places the biomolecule under structural
tension, thereby incurring an energetic penalty.
This can be clearly seen in the interaction between the
TATA box binding protein and DNA (figure \ref{fig:amino}, bottom left),
which forces the DNA to adopt a highly kinked structure.

As biomolecules are flexible and
change shape significantly due to thermal fluctuations, the entropic
term $T\Delta S$ is also important in the free energy change $\Delta G$.
In general, protein folding leads to a
reduction in entropy as the unravelled polypeptide chain is constrained
into its folded structure. Biomolecular association is often (but not always) accompanied by a reduction in entropy, because two previously
independent molecules combine into a single complex, and because
accommodating another molecule often inhibits the conformational
flexibility of the participants. In addition, the solvent participates
in mediating biochemical interactions. A number of the amino acids are
hydrophobic. These hydrophobic residues are usually confined to the
centre of the protein during folding, however, evolution has designed
proteins which contain hydrophobic binding pockets which promote the
binding of other hydrophobic molecules. The overall stability of the
protein or biomacromolecular complex depends on the sum of all of these
different competing terms.

Arguably, the most impressive macromolecular structures that have evolved
act as molecular motors. Two examples of molecular machines are shown
figure \ref{fig:amino}. DNA helicase (bottom centre), which separates
two strands of DNA, is one of the simplest molecular machines. The
ribosome (bottom right) is far larger due to its more complex function
of translating the amino acid code into protein. Biological motors
convert chemical energy into mechanical work (or visa versa) by
amplifying localised changes in chemistry into large changes in
global conformation.

\section{Computer modelling of Biomolecules}
\label{sec:biosim}

The most common technique for obtaining
dynamical information at the atomistic level for biomolecules
is molecular dynamics (MD) simulation.
MD provides the positions
and velocities of all of the atoms in the system as a function of time
through a numerical integration of Newton's equations using a very short
time step $dt$ (generally $2 fs$).  This is necessary to ensure
stability of the numerical approximation and to capture the high
frequency vibrations of individual bonds.
The pair-wise forces acting on
each atom are calculated from the gradient of an appropriate potential
energy function which is known as the force-field. This force-field uses a harmonic potential to describe covalent bonds, electrostatic interactions are calculated using Coulomb's law based on a set of empirical partial charges assigned to every atom, and dispersion is modelled using the van der Waals potential. The accuracy of an MD
simulation depends critically on an appropriate choice of the
force-field parameters
used in this potential energy function.
These empirical parameters are derived from
quantum mechanical calculations on molecular fragments,
and are continuously under revision as the computational methods
for obtaining them improve.

The environment of a biomolecule is extremely important. The most
accurate MD simulations incorporate a
large periodic box of water to mimic solution conditions, while simulations of
membrane proteins embed the protein in lipid molecules, as shown in
figure \ref{fig:amino} (top left).
The simulation of highly charged molecules, such as DNA, requires
particularly careful treatment of long-range electrostatic interactions.
Most biomolecular simulation codes implement the Ewald summation
technique,
which is achieved computationally
using a Fast Fourier Transform (FFT).
Despite recent advances
in the efficiency and parallelisability of FFT algorithms, this portion of
the force-field calculation is still costly.
For a more detailed description of
modelling methods applied to biomolecules, see \cite{Leach}.

Given an experimentally derived structure of a drug/protein complex,
atomistic simulation can provide a good estimate for the interaction
energy stabilising the complex.
However, as it is not computationally possible to explore the full
conformational space accessible to the reactants and the products,
the entropic portion of the free energy change cannot be calculated
with any certainty.
Hence, we are unable to accurately predict binding free energies.
Pharmaceutical scientists would like to be able to predict the
binding affinity of a whole combinatorial library of potential
new drugs \emph{in silico} before undertaking the expensive process
of chemical synthesis, protein expression and biochemical
measurements of binding free energies.
Numerical studies are still widely employed in the
pharmaceutical industry to test the potential of new drug molecules
but the approximations employed,
such as rigid molecules, mean the results are not very reliable.
The study of non-equilibrium processes, such as protein folding,
are even more computationally demanding, and
the conformational changes associated with the action of
a molecular machine such as RNA polymerase,
lie way beyond our predictive abilities.

The maximum achievable simulation timescale for a (relatively small)
protein containing 162 amino acids in solution ($\sim 3\times 10^4$
total atoms) is around 10$\mu$s \citep{proteinfold}. Recent benchmarking studies using the
AMBER molecular dynamics software indicate that it is possible to obtain
around 100 ns a day on 32 CPU's for a small biomolecule,
so a 10$\mu$s simulation on 32 processors requires $\sim$3200 CPU days.
Larger biomolecular complexes are prohibitively expensive for even
nanosecond timescales: e.g.~a 20 ns simulation of the ribosome, with
circa 2.6 million atoms, required $\sim 10^6$ CPU hours on 768 CPUs in
2006 \citep{ribosome}, which corresponds to $\sim$55 days of CPU.
The ribosome synthesises new proteins at a rate of 1 amino acid every 0.1s
\citep{ribotime},
so to capture the action of such a molecular motor 
would require simulation timescales $\sim 10^6$ times longer,
which corresponds to $10^{12}$ CPU hours, or almost 1.5 million years.
Using specialised hardware,
MD simulations of 1ms in just over 2 months are predicted for the small
protein dihydrofolate reductase ($\sim$25,000 atoms including solvent)
\citep{anton}, faster by a factor of over 100.
This speed up will allow significant progress, but a further factor
of $10^3\sim 10^4$ is needed to make real breakthroughs.
The prospect of using quantum computation as a
tool in molecular biology is thus very attractive, if it can deliver
the necessary improvements in capabilities.

Before discussing how a quantum computer could simulate a complex
system like a protein interacting with a drug,
or even an entire cell, it is worth considering the nature of computer
simulation and what we achieve by its use.
Essentially, we are testing our most accurate models of the
real world: by calculating in detail what they predict, and comparing
this with our observations
(for example, an experimentally determined binding constant for a
protein-drug interaction).
If our calculations and observations agree as well as we anticipate, this is
evidence our models are appropriate and that we understand (at some level)
how the system works.
We may then use our computer simulations to
predict things we haven't yet observed, or provide more details of
processes that are hard to obtain by experiment.

That computation of any sort works in a useful way is not trivial.  For
some systems, we can calculate things much faster than it takes to do
the experiment (the trajectory for sending a space probe to Saturn, for
example). For biomolecules, it takes far longer to run the simulation than
the real system takes to do the same thing.  This is because
we have a simple model of Newtonian gravitation that works
extremely well for satellites and planets, even though the model does
not have analytic solutions for more than two bodies.  While a
planet is much more complex than a protein, most of this detail is
irrelevant for how a space probe travels round it in orbit,
so is not included in the simulations.  Biomolecules are
costly to simulate because far more of the details contribute to the
behaviour of the system.

We do have a simple quantum-mechanical model of electrons,
protons and neutrons, and how they behave when clumped
together as atoms and molecules.
Quantum effects are integral to chemical reactions
when covalent bonds are broken and reformed,
for example, during enzyme catalysis.
However, in biology, most processes involve more
subtle energetic changes, where quantum effects can be well-approximated
using force-field parameters derived from quantum chemistry simulations,
combined with classical mechanics.
Notable exceptions include charge and
energy transport in photosynthesis, where the key reaction takes place
over a few tens of atoms \citep{mohseni08a,plenio08a,sarovar09a},
and highly sensitive receptors for light that can
distinguish polarisation \citep[e.g.,][]{roberts09a}
or detect single photons \citep[e.g.,][]{lillywhite77a}.
Ironically, quantum processes such as these are better understood
-- because of the small scales over which they take place they are
accessible using today's computers -- than the largely classical
processes governing the cells in which they take place.

\section{Quantum computing applied to biomolecules}
\label{sec:bioqc}

It is now twenty five years since \cite{feynman82a} and
\cite{deutsch85a} first proposed that quantum systems should
be able to process information fundamentally more efficiently
than classical computers.  They both (independently) perceived
that a superposition of multiple quantum trajectories looks like
a classical parallel computer, which calculates the result
of many different input values in the time it takes for one processor
to do one input value.  In quantum systems, this parallel
processing comes ``for free'' with the superposition of
the quantum state, and promises an exponential saving in the memory
and processing time required for suitable problems.
Simulation of quantum systems was the original idea from
\citeauthor{feynman82a} for what a quantum computer could do better than
classical, and this is expected to be one of the first useful
applications of small quantum computers.
For more detailed discussion, see \cite{kendon10a}. 
Less work has been done on how to apply quantum computers to
classical simulations, where limitations in classical computing power
show up just as keenly, as our discussion of biomolecular simulation
makes clear.  \cite{harrow08a} provide a quantum algorithm for solving linear
systems of equations, while quantum lattice gas methods
\citep{boghosian96a,meyer01a} can be used for classical as well as
quantum simulations.
Solving eigenvalue equations \citep{abrams99a} can be done exponentially
more efficiently too.

In quantum chemistry it actually helps to keep a
more detailed model for quantum simulations \citep{kassal08a},
the approximations used in classical computations would slow the
quantum computer down.
Calculations at the quantum chemical level provide
the empirical parameters essential to construct
the force-fields for biomolecular simulations,
one important area where quantum computers can contribute \citep{fan09a}.
This improves accuracy, but does not increase speed.
However, it won't help us to keep the quantum details for biomolecular
simulations of systems with hundreds of thousands of atoms.
By using classical dynamical models we've already made an exponential
saving in resources by reducing the state space from Hilbert space
to classical degrees of freedom.  What we need are more efficient ways to
perform calculations of classical molecular dynamical systems.
A quantum computer can offer a significant advantage only if
we can employ a quantum algorithm with a better scaling than offered by
the classical methods.  Even a small polynomial (quadratic or
even less) advantage would be significant in practice, given the
large size of the systems we wish to simulate.
There are three main ways in which a quantum computer could provide
an improvement.

\subsection{Encode the system in a quantum superposition}

This would use quantum parallelism to
mimic the way current classical parallel algorithms work.
A single classical computation would be carried out using
exponentially less memory, at the end of which we obtain only
an exponetially small amount of the full information.
This is thus suitable only for problems where the result
is a global average of some sort that can be efficiently
extracted from the final state.
It requires the whole computation to be quantum from
start to finish, requiring quantum coherence to be maintained
for all of a large, long computation.

\subsection{Perform multiple computations in superposition}

The system is encoded in the same way as for classical
simulations, i.e., no saving in memory, but the quantum
superposition allows us to perform several computations in parallel.
A single simulation could
process several different initial states at the same time,
or all branches of a section of the computation could be
calculated simultaneously.
This approach requires some sort of quantum trick
to select for favored outcomes at the expense
(destructive interference of) less-favored outcomes.
It could be applied to protein folding; for example,
here many possible configurations can be explored, but
those with higher energy are penalised.  Energy minimisation
problems can be mapped onto adiabatic quantum computation
see, for example, \cite{perdomo08a}, for discussion of how
to do this with the hydrophobic-polar model for protein-folding.
The potential savings here depend on how many different input
states, or paths, need to be calculated to find the right one.
An exponentially large superposition is possible, if
required by the problem.

Multiple simultaneous computations could also
be used to explore the phase space sufficiently to
provide an estimate of the entropic term in the free
energy.  This would not increase the size of the system
that can be simulated, but would provide a really important
improvement in the accuracy of the calculations, since
the entropic term is neglected or poorly estimated using
current methods.  This is thus a very attractive option for
biomolecular MD simulations.

\subsection{Quantum subroutines}

A hybrid strategy in which costly parts of the
computation of a single time step are turned into more
efficient quantum subroutines.  This has the advantage
of requiring quantum coherence only for
shorter time scales, so may be feasible sooner than fully
quantum methods.  There is no saving in memory, and
the running time is reduced by a constant factor determined
by the per time step speed up.
This method is appropriate for dynamical studies where we need all
the classical information.

The most obvious subroutine to quantumise, the Fourier
transform, will not yield the speed up one would naively expect.
Although the quantum Fourier transform (QFT) plays a key role
in quantum algorithms with an exponential speed up over the best known
classical, e.g., the factoring algorithm due to \cite{shor95a},
it has been shown to be efficiently classically simulatable
when either the input is a classical (separable) state,
or the output is measured immediately following the QFT
\citep{browne06a,aharonov06a}.  Both of these conditions apply here.
However, a less dramatic speed up is possible, if
the quantum FT processor takes less time than the classical FT, and
the data can be efficiently copied between the quantum and classical
processors.  The FFT scales as $O(N\log N)$ while the QFT is $O(N)$,
but what matters is the actual time taken, rather than the
scaling, which we won't be able to determine without
details of the quantum processor architecture.

This method could be applied to protein folding, or
other dynamics where we have to explore a number of paths
to find the optimum choice.  In this case, we use a quantum
subroutine to find the best move at each step, then
classically implement the single chosen move.
Unlike in option 2., the dynamics are obtained step by step.

\section{Future perspectives}
\label{sec:future}

Biomolecular simulation is extremely computationally demanding,
and there are no ``free lunches'' when it comes to processing
large complex systems.  Nonetheless, quantum computers have
special capabilities that could make a real difference to
our ability to compute and predict the properties of
biological systems at the cellular level.
Most importantly, it is the ability of a quantum computer
to explore many classical paths simultaneously that offers
a potential method to overcome the problem of finding the
minimum \emph{free energy}, as opposed to just the
minimum energy, in an optimisation problem such as
protein folding or molecular docking.
State of the art MD simulations now routinely include ``repeat''
simulations, in which a number of initial conformations are
investigated in parallel to check the robustness of any
conclusions against thermal noise.
The potentially massive parallelisation provided by a
quantum computer offers the possibility of evolving a
whole thermodynamic ensemble, rather than a single trajectory.
Quantumised ensemble dynamics algorithms may one day make
calculating the binding free energy of a complex as routine as
current calculations of the binding energy. With a quantum computer,
it may be as straightforward to calculate the optimal chemistry
of the drug necessary to turn off an cancerous gene, for example,
as it now is to calculate the precise trajectory required for a
rocket to reach the moon.

\begin{acknowledgements}
We thank Katie Barr for help with the references,
and Binbin Liu for providing the membrane protein co-ordinates.
VK funded by a Royal Society University Research Fellowship.
\end{acknowledgements}

\small

\normalsize

\label{lastpage}
\end{document}